\documentclass[twocolumn,showpacs,amsmath,amssymb]{revtex4}
\usepackage{graphicx}

\begin{document}

\title{Tensor Network and Black Hole}
\author{Hiroaki Matsueda${}^{a}$\footnote{matsueda@sendai-nct.ac.jp}}
\author{Masafumi Ishihara${}^{b}$\footnote{masafumi@wpi-aimr.tohoku.ac.jp}}
\author{Yoichiro Hashizume${}^{c}$\footnote{hashizume@rs.tus.ac.jp}}
\affiliation{
${}^{a}$Sendai National College of Technology, Sendai 989-3128, Japan \\
${}^{b}$WPI-Advanced Institute for Materials Research (WPI-AIMR), Tohoku University, Sendai 980-8577, Japan \\
${}^{b}$Department of Physics, Tokyo University of Science, Tokyo 102-0073, Japan
}
\date{\today}
\begin{abstract}
A tensor network formalism of thermofield dynamics is introduced. The formalism relates the original Hilbert space with its tilde space by a product of two copies of a tensor network. Then, their interface becomes an event horizon, and the logarithm of the tensor rank corresponds to the black hole entropy. Eventually, multiscale entanglement renormalization anzats (MERA) reproduces an AdS black hole at finite temperature. Our finding shows rich functionalities of MERA as efficient graphical representation of AdS/CFT correspondence.
\end{abstract}
\maketitle

\section{Introduction}

Applications of anti-de Sitter space / conformal field theory (AdS/CFT) correspondence~\cite{Maldacena} to statistical and condensed matter physics are hot topics in string theory. On the other hand, it is recognized that a new class of variational anzats in statistical physics, so called multiscale entanglement renormalization anzats (MERA), would be a discrete version of the AdS/CFT correspondence~\cite{Vidal,Vidal2,Pfeifer,Evenbly,Swingle,Matsueda}. Both of them overcomes difficulties of real-space renormalization in critical systems. It is thus attracting attention to examine their complementarity in a mathematical level.

The MERA is a kind of tensor network (product) states (TNS, TPS) of quantum many-body systems on lattices~\cite{Matsueda,Verstraete,Vidal3,Verstraete2}. Historically, the TNS formalism was constructed so that the ground-state variational wave function satisfies the entanglement-entropy scaling. The examination was first developed for gapped cases. In the gapped cases, the entropy obeys the well-known area-law scaling~\cite{Bekenstein,Hawking,Hawking2,Bombelli,Srednicki}. In spatially one dimension (1D), the wave function appropriate for the scaling is matrix product state (MPS). Actually, the MPS is numerically optimized by the density matrix renormalization group (DMRG) method, and the DMRG is known to be the most powerful method in quantum 1D systems~\cite{White,Ostlund,Verstraete3}. A natural generalizartion of the MPS to higher dimensions is to make the tensor contraction that represents a set of short-range entangled pairs on bonds. In that sense, the TNS is also called projected entangled pair state (PEPS). When the tensor rank and the surrounding area of a partial system are $m$ and $A$ respectively, the entanglement entropy between the partial system and its enviromnent is given by $S_{EE}=A\ln m\propto A$. On the other hand, in critical cases, the area-law scaling violates logarithmically~\cite{Holzhey,Calabrese,Calabrese2,Plenio,Riera,Vidal4,Wolf,Gioev,Gioev2,Li,Barthel}. Then, we must take a large $m$ value of the order of the total sites $L$, if we keep the basic TNS structure. However, it is possible to construct a hierarchical tensor network in one higher dimension that has a managable size of the tensor rank and shows the logarithmic divergence of the entropy. That is a concept of the MERA network.

\begin{figure}[htbp]
\begin{center}
\includegraphics[width=7.5cm]{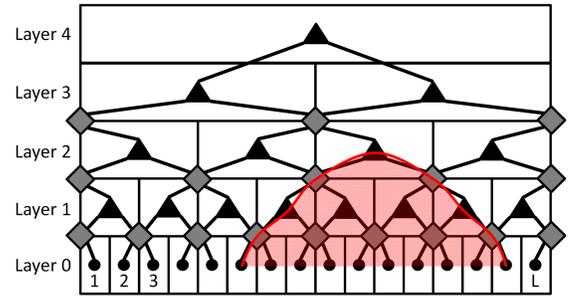}
\end{center}
\caption{2D binary MERA network. Filled dots, diamonds, and filled triangles are the original sites in a quantum 1D critical system, disentangler tensors, and isometries, respectivery. The vertical direction denotes renormalization flow. A red line represents the causal cone enclosing a partial system. The surface area of the causal cone is given by the sum of the number of the boundary points in each layer.}
\label{fig1}
\end{figure}

The MERA network has its graphical representation as shown in Fig.~\ref{fig1}, and it is easy to visualize the entanglement structure inherent in the network. A key ingredient of MERA is that the network is composed of layers, each of which has its own energy and length scales. The layered structure emerges as a result of repeating the block-spin transformation as well as disentangling transformation. In an another viewpoint, the tensors with large $\chi$ in the TNS is decomposed into a set of tensors with smaller dimensions and with different functionalities. Then, the total entanglement entropy is given by the sum of the entropy in each layer compatible with the area-law scaling. The reason for the appearance of the area law is that the smaller tensor rank represents more classical-like short-range correlation in this extended space. Graphically, the sum corresponds to the outside area of the causal cone in the discrete AdS space, and actually this is comparable to the Ryu-Takayanagi formula~\cite{Ryu}.

In this paper, we focus on general TNS and in particular MERA at finite temperature. In comparison with the ground states, finite temperature properties of the TNS are less understood. Because of the potential complementarity of MERA and AdS/CFT, the understanding also gives us deeper insight for application of the AdS/CFT to condensed matter physics~\cite{Liu,Ogawa,Huijse,Shaghoulian}. In the application, it is necessary to deform the asymptotically AdS metric so that the IR geometry has a black hole solution. The black hole is a source of coarse graining, and determines the temperature at the conformal boundary of the AdS space. Then, we would like to know whether the black hole naturally emerges from the TNS formalizm without any phenomenological assumptions. This is the purpose of this paper.

In a viewpoint of the CFT, the TNS has rich functionalities. Let us briefly look at a basic property of the MPS that is the most primitive TNS in 1D. The property is about a dicrete version of the Calabrese-Cardy formula for the entanglement entropy~\cite{Calabrese,Calabrese2}, the so called scaling of entanglement support given by $S_{EE}=(c\kappa/6)\ln\chi$, where $c$ is the central charge, $\kappa=6/c(\sqrt{12/c}+1)$ is the finite-entanglement scaling exponent, and $\chi$ is the matrix dimension~\cite{Tagliacozzo,Pollmann,Matsueda2}. It has been shown that the correlation length $\xi$ is given by $\xi=\chi^{\kappa}$. This means that the MPS is a very simple structure but catches the essential feature of the CFT. Therefore, we expect that the black hole is also described by the TNS. We will confirm this statement with use of the MERA network.

Since the MERA is a wave-function approach, it is straightforward to use thermofield dynamics (TFD) for finite-temperature formalism of quantum field theory (the readers may be aware of equivalence between TFD and dual CFT in a special case)~\cite{Takahashi,Israel,Maldacena2,Cantcheff,Czech,Raamsdonk}. For our purpose, we reformulate the original TFD so that the TFD wave function matches well with the tensor network representation. We will find that by this reformulation the event horizon naturally appears as a result of hidden quantum entanglement between the original Hilbert space and its tilde space. The entanglement produces a combined tensor network, and we will examine the basic properties of the network.

The paper is organized as follows. Section II is the main part of this paper. We first examine a single-site model, and find a method for constructing the TNS formalism of the thermal state. Then, we examine the finite-temperature MERA network. We will present microscopic derivation of the black hole entropy, the event horizon, and the temperature at the conformal boundary of the AdS space. In Sec. III and Sec. IV, we discuss related topics and summarize our study.

\section{Tensor Network Representation of Thermal State}

\subsection{Entanglement between Dual Hilbert Spaces and Thermal Entropy}

Let us start with a particular model in TFD. In addition to the original Hilbert space, we introduce the tilde space that is isomorphic and decoupled to the original space. Then, it is able to represent a thermal average by the expectation value of the thermal state. We define a generator $G$ and its evolution operator $U(\theta)$ by
\begin{eqnarray}
G&=&i\hbar\omega\left(a\tilde{a}-a^{\dagger}\tilde{a}^{\dagger}\right), \\
U(\theta)&=&e^{i\theta G},
\end{eqnarray}
where the parameters $\theta=\beta/2=1/2k_{B}T$ and $\omega$ characterize inverse temperature and the energy scale of excitation modes, respectively. The Fermion operators $a$ and $\tilde{a}$ satisfy the anti-commutation relations
\begin{eqnarray}
&& \left\{ a,a^{\dagger}\right\} = \left\{ \tilde{a},\tilde{a}^{\dagger}\right\} = 1 , \\
&& \left\{ a,a\right\} = \left\{ \tilde{a},\tilde{a}\right\} = 0 .
\end{eqnarray}
Here, we assume
\begin{eqnarray}
\left\{ \tilde{a},a\right\} = \left\{ \tilde{a},a^{\dagger}\right\} = 0 .
\end{eqnarray}
In the following, we define the vacuum state by $\left|0\tilde{0}\right>=\left|0\right>\otimes\left|\tilde{0}\right>$, $a^{\dagger}\left|0\right>=0$, and $\tilde{a}^{\dagger}\left|\tilde{0}\right>=0$. We use the notation
\begin{eqnarray}
a^{\dagger}\left|0\tilde{0}\right> &=& \left|1\tilde{0}\right> , \\
\tilde{a}^{\dagger}\left|0\tilde{0}\right> &=& \left|0\tilde{1}\right> , \\
a^{\dagger}\tilde{a}^{\dagger}\left|0\tilde{0}\right> &=& \left|1\tilde{1}\right>.
\end{eqnarray}

Let us consider the following thermal state
\begin{eqnarray}
\left|O(\theta)\right>&=&U(\theta)\left|0\tilde{0}\right> \\
&=&\left(u(\theta)+v(\theta)a^{\dagger}\tilde{a}^{\dagger}\right)\left|0\tilde{0}\right> \label{wf},
\end{eqnarray}
where we can find
\begin{eqnarray}
u(\theta)&=& \cos\left(\hbar\omega\theta\right) , \\
v(\theta)&=& \sin\left(\hbar\omega\theta\right) .
\end{eqnarray}
In order to examine the entanglement entropy of this system, we define the partial density matrix $\rho$ by
\begin{eqnarray}
\rho&=&\tilde{{\rm Tr}}\left|O(\theta)\right>\left<O(\theta)\right| \\
&=&u(\theta)^{2}\left|0\right>\left<0\right|+v(\theta)^{2}\left|1\right>\left<1\right| , \label{dm}
\end{eqnarray}
where $\tilde{{\rm Tr}}$ traces over degrees of freedom in the tilde space. In a context of TFD, this is nothing but the thermal density matrix. At the same time, $\rho$ represents the amount of entanglement between the original and the tilde Hilbert spaces. The entanglement entropy is given by
\begin{eqnarray}
S_{EE}&=&-{\rm Tr}\bigl(\rho\ln\rho\bigr) \\
&=&-u(\theta)^{2}\ln u(\theta)^{2}-v(\theta)^{2}\ln v(\theta)^{2} \\
&\le& \ln 2.
\end{eqnarray}
On the other hand, the thermal entropy of this system is clearly (the system takes $\left|0\right>$ or $\left|1\right>$)
\begin{eqnarray}
S_{T}=k_{B}\ln 2.
\end{eqnarray}
Identifying $S_{EE}$ with $S_{T}/k_{B}$~\cite{Bombelli,Srednicki,Fiola,Susskind,Hawking3,Emparan,Solodukhin}, we find $\hbar\omega\theta=\pi/4$ and
\begin{eqnarray}
T=\frac{2\hbar\omega}{\pi k_{B}} , \label{Hawking}
\end{eqnarray}
which maximizes $S_{EE}$ and then $S_{EE}$ looks like $S_{T}$.

\subsection{Vector Product Form of Thermal State}

In order to examine the physical meaning of Eq.~(\ref{Hawking}), we first transform Eq.~(\ref{wf}) into a vector product form. That is represented as
\begin{eqnarray}
\left|O(\theta)\right>=\sum_{m,\tilde{n}=0,1}A^{m}A^{\tilde{n}}\left|m\tilde{n}\right>, \label{tildeTNS}
\end{eqnarray}
where $A^{m}$ and $A^{\tilde{n}}$ are $\chi$-dimensional vector and its tilde conjugate, respectively. The index $m$ ($\tilde{n}$) takes $0$ or $1$. Usually it is considered that the original and the tilde Hilbert spaces are independent. However, this hidden correlation mediated by the rank $\chi$ plays an important role on the emergence of the black hole. The tilde conjugate of the vector is not defined in the original TFD, but in this paper we define it by the transposition as well as the complex conjugate of the vector elements. This wave function is nothing but the MPS. Here we consider the single-site problem with open boundary condition, and then the edge matrix is terminated by an appropriate vector in order to get a scalar coefficient.

We introduce the following representation
\begin{eqnarray}
A^{m}=\left(A_{1}^{m}, A_{2}^{m}, \cdots, A_{\chi}^{m}\right) , 
A^{\tilde{n}}=\left(\begin{array}{c}
A_{1}^{\tilde{n}} \\ A_{2}^{\tilde{n}} \\ \vdots \\ A_{\chi}^{\tilde{n}}
\end{array}\right) .
\end{eqnarray}
Then, we obtain the following solutions
\begin{eqnarray}
A^{0}=\left(\sqrt{u(\theta)}, 0\right) , 
A^{1}=\left(0, \sqrt{v(\theta)}\right) , \label{sol1}
\end{eqnarray}
and
\begin{eqnarray}
A^{\tilde{0}}=\left(\begin{array}{c}
\sqrt{u(\theta)^{\ast}} \\ 0
\end{array}\right) , 
A^{\tilde{1}}=\left(\begin{array}{c}
0 \\ \sqrt{v(\theta)^{\ast}}
\end{array}\right) . \label{sol2}
\end{eqnarray}
We can find more general solutions, but according to symmetry of two Hilbert spaces, we take the vector elements so that $A^{\tilde{n}}$ becomes a copy of $A^{m}$. In the present case, the proper (minimal) $\chi$ value is $2$. When we compare Eq.~(\ref{dm}) with Eqs.~(\ref{sol1}) and (\ref{sol2}), we clearly find that the thermal entropy counts the degrees of freedom at the interface between $A^{m}$ and $A^{\tilde{n}}$ when they are maximally entangled. Futhermore, there is a fact that the entropy of extremal black holes known so far has been explained by assuming maximally entangled states, and this fact also matches with our situation. In that sense, our entropy seems to behave as the black hole entropy and the event horizon is located at the interface of $A^{m}$ and $A^{\tilde{n}}$. Then, Eq.~(\ref{Hawking}) corresponds to the Hawking temperature of our system. The maximally entangled system appears in the equal probability case ($p_{i}=1/\chi$ for $i=1,2,...,\chi$) even when the degree of freedom increases. Then, the upper bound of the entanglement entropy is given by 
\begin{eqnarray}
S_{EE}^{max}=-\sum_{i=1}^{\chi}p_{i}\ln p_{i}=\ln\chi, \label{entangle}
\end{eqnarray}
and this also leads to the thermal entropy.

In general, the thermal state is represented as
\begin{eqnarray}
\left|\Psi\right>=\rho^{1/2}\left|I\right> ,
\end{eqnarray}
where $\rho$ is the density matrix, and we take $\left|I\right>=\sum_{i}\left|i\tilde{i}\right>$ according to the general representation theorem~\cite{Suzuki}. We decompose $\rho$ into singular values $\{\lambda_{j}\}$ as
\begin{eqnarray}
\rho^{1/2}=V\Lambda\tilde{V} ,
\end{eqnarray}
where $V$ and $\tilde{V}$ are column unitary matrices, $\Lambda$ is a diagonal matrix and ${\rm diag} \Lambda = (\lambda_{1},\lambda_{2},...)$. Then the $\chi$ value is given by
\begin{eqnarray}
\chi={\rm rank}(V\Lambda^{1/2})={\rm rank}(\Lambda^{1/2}\tilde{V}) .
\end{eqnarray}

\subsection{Decomposition of Vector into Truncated MERA Network and Emergence of AdS Black Hole}

For general $L$-sites systems, Eq.~(\ref{tildeTNS}) is extended as
\begin{eqnarray}
\left|\Psi\right> &=& \sum_{\{m_{j}\}}\sum_{\{\tilde{n}_{j}\}}A^{m_{1}m_{2}\cdots m_{L}}A^{\tilde{n}_{1}\tilde{n}_{2}\cdots\tilde{n}_{L}} \nonumber \\
&& \times\left|m_{1}m_{2}\cdots m_{L}\tilde{n}_{1}\tilde{n}_{2}\cdots\tilde{n}_{L}\right>, \label{tildeTNS2}
\end{eqnarray}
where $m_{j}$ and $\tilde{n}_{j}$ are local valuables and
\begin{eqnarray}
A^{m_{1}m_{2}\cdots m_{L}}A^{\tilde{n}_{1}\tilde{n}_{2}\cdots\tilde{n}_{L}}=\sum_{\alpha=1}^{\chi}A_{\alpha}^{m_{1}m_{2}\cdots m_{L}}A_{\alpha}^{\tilde{n}_{1}\tilde{n}_{2}\cdots\tilde{n}_{L}} .
\end{eqnarray}
It is argued in ref.~\cite{Javier,Javier2} that truncating some tensors from the complete MERA network roughly represents the AdS black hole. We derive it from a microscopic viewpoint. In our notation, $A_{\alpha}^{m_{1}m_{2}\cdots m_{L}}$ corresponds to the truncated tensor network. Thus, we decompose it into a set of tensors with smaller dimensions. The decomposition of $A_{\alpha}^{m_{1}m_{2}\cdots m_{L}}$ into any network is always possible. Depending on criticality of our target models, we select an appropriate network. The tensor geometry should match with the symmetry of the original quantum system.

\begin{figure}[htbp]
\begin{center}
\includegraphics[width=7.5cm]{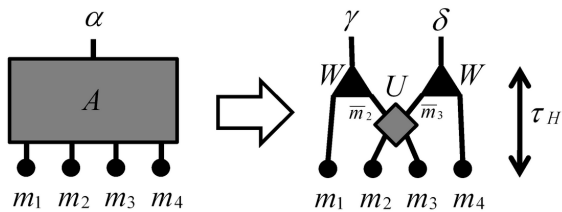}
\end{center}
\caption{Decomposition of the vector $A$ into a set of tensors.}
\label{fig2}
\end{figure}

Hereafter we focus on the conformal field theory described by the 2D MERA network. Then, we decompose $A_{\alpha}^{m_{1}m_{2}\cdots m_{L}}$ into a MERA-like network. It is noted that the presence of the index $\alpha$ leads to the incomplete MERA network. Figure~\ref{fig2} is an example that is the four-sites MERA network under the open boundary condition but the top tensor is truncated (the network is terminated). This diagram is mathematically represented as
\begin{eqnarray}
A_{\alpha}^{m_{1}m_{2}m_{3}m_{4}}=\sum_{\bar{m}_{2},\bar{m}_{3}}W_{m_{1}\bar{m}_{2}}^{\gamma}U_{m_{2}m_{3}}^{\bar{m}_{2}\bar{m}_{3}}W_{\bar{m}_{3}m_{4}}^{\delta} ,
\end{eqnarray}
with use of isometry $W$ and disentangler $U$. Since the vector rank is $\chi$, the indices $\gamma$ and $\delta$ take $1,2,...,\sqrt{\chi}$.

We generalize the above result. The total degrees of freedom at the interface is originally $\chi$, but let us assume that the tensor rank and the number of the tensors at the interface are respectively $m$ and $A$ after the decomposition. Then, $\chi$ is represented as
\begin{eqnarray}
\chi=m^{A}. \label{cond1}
\end{eqnarray}
Here, $A$ is related with $L$ by the following condition
\begin{eqnarray}
\frac{L}{\eta^{\tau_{H}}}=A, \label{cond2}
\end{eqnarray}
where $\tau_{H}$ is the layer number starting from $0$. In the binary MERA network, we take $\eta=2$. In general, each renormalization process merges $\eta$-sites together. Substituting Eqs.~(\ref{cond1}) and (\ref{cond2}) into the upper bound of the entanglement entropy, we obtain
\begin{eqnarray}
S_{EE}^{max} = \frac{L}{\eta^{\tau_{H}}}\ln m. \label{cond3}
\end{eqnarray}
Now our discrete AdS space in Fig.~\ref{fig1} is represented by the metric
\begin{eqnarray}
ds^{2}=\bigl\{d\bigl(\tau\ln\eta\bigr)\bigr\}^{2} + \bigl(\eta^{-\tau}dx\bigr)^{2},
\end{eqnarray}
and changing the valuable $\tau$ to $z=\eta^{\tau}$ ($z$ is normalized by the AdS curvature), we obtain the standard notation
\begin{eqnarray}
ds^{2}=\frac{dz^{2}+dx^{2}}{z^{2}}.
\end{eqnarray}
Thus, Eq.~(\ref{cond3}) is transformed into
\begin{eqnarray}
S_{EE}^{max} = \frac{L}{z_{H}}\ln m, \label{cond4}
\end{eqnarray}
where $z_{H}=\eta^{\tau_{H}}$. According to the CFT at finite temperature, the entanglement entropy of the original 1D quantum systems is given by
\begin{eqnarray}
S_{EE}=\frac{c}{3}\ln\left(\frac{\beta}{\pi\epsilon}\sinh\left(\frac{\pi L}{\beta}\right)\right) , \label{cond44}
\end{eqnarray}
where $\epsilon$ is a UV cutoff. Assuming that $\beta$ is small enough, we expand Eq.~(\ref{cond44}) as
\begin{eqnarray}
S_{EE}\simeq \frac{c}{3}\ln\left(\frac{\beta}{2\pi\epsilon}\right) + \frac{c}{3}\frac{\pi L}{\beta}. \label{cond5}
\end{eqnarray}
The second term is proportional to the system size $L$, and thus $S_{EE}$ obeys the volume law at high temperature. When we identify Eq.~(\ref{cond4}) with Eq.~(\ref{cond5}), we find that
\begin{eqnarray}
k_{B}T=\left(\frac{3}{c\pi}\ln m\right)\frac{1}{z_{H}}\propto z_{H}^{-1}. \label{cond6}
\end{eqnarray}
This is nothing but the temperature scale arizing from the AdS black hole, and is consistent with the previous results based on the holographic principle~\cite{Liu,Ogawa,Huijse,Shaghoulian,Swingle2}. Our case is similar to ref.~\cite{Azeyanagi} in a sense that the black hole entropy is interpreted as the entanglement entropy in CFT living on the boundary of the AdS space. In our formulation, the separation of the CFT and the event horizon was derived from decomposition of $A_{\alpha}^{m_{1}m_{2}\cdots m_{L}}$ into the truncated MERA network.

Before going to the next subsection, we further comment on the physical meaning of Eq.~(\ref{cond6}). We require consistency between our result and the holographic theories up to the coefficient. Then, the coefficient of $z_{H}^{-1}$ should be given by
\begin{eqnarray}
\frac{3}{c\pi}\ln m = \frac{1}{2\pi}, \label{cond7}
\end{eqnarray}
and this requires that
\begin{eqnarray}
m=e^{c/6}\sim 1 . \label{cond8}
\end{eqnarray}
Here we have supposed the minimal series ($0<c<1$) and the Gaussian CFT ($c=1$). The right hand side of Eq.~(\ref{cond7}) has been obtained for continuous systems. Thus, Eq.~(\ref{cond8}) is just a criterion of the $m$ value. However, the result indicates that the MERA generates a really 'classical' space near the event horizon, since the quantum entanglement has been almost vanished for the $m$ value. This small $m$ value might be the reason for the success of semiclassical treatment of the black hole thermodynamics.

\subsection{MERA - Tilde MERA Combined Network}

\begin{figure}[htbp]
\begin{center}
\includegraphics[width=7cm]{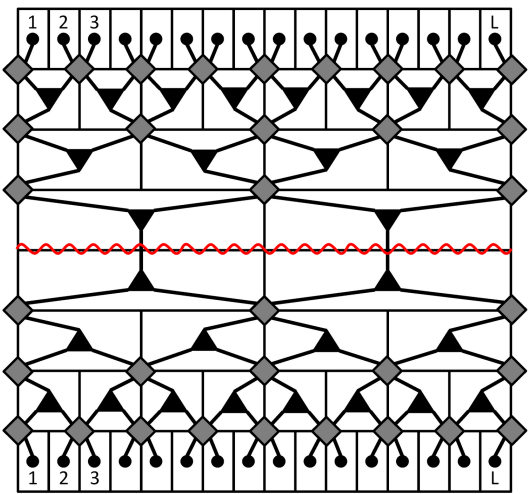}
\end{center}
\caption{Combination of MERA (lower half) and tilde MERA (upper half) networks. A red wavy line represents an event horizon. Two pairs of isometries across the horizon are entangled.}
\label{fig3}
\end{figure}

\begin{figure}[htbp]
\begin{center}
\includegraphics[width=7cm]{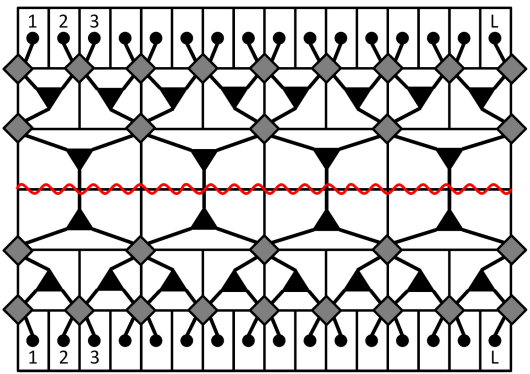}
\end{center}
\caption{Higher temperature case of MERA - tilde MERA combined network. When we fix system size $L$, larger $\chi$ leads to the smaller layer number.}
\label{fig4}
\end{figure}

Up to now, we have considered a method for decomposition of $A_{\alpha}^{m_{1}m_{2}\cdots m_{L}}$ into a terminated MERA network. Let us imagine that the coefficient of the basis in Eq.~(\ref{tildeTNS2}) is a product of two terminated MERA networks. In the tilde-space side, we call it as 'tilde MERA' network. Figure~\ref{fig3} shows schematic representation of our extended MERA network. We also show higher temeperature case in Fig.~\ref{fig4}. The total system is composed of the original and tilde MERA networks. At finite temperature, some upper layers of the original MERA network is truncated as already discussed. Two truncated MERA networks are pasted together at the position of the event horizon. The surface area of the horizon is determined by Eq.~(\ref{cond2}), and the temperature of the conformal boundary is given by Eq.~(\ref{cond6}). In ref.~\cite{Cantcheff}, a similar viewgraph is introduced in a context of gravitational collapse in the AdS spacetime. Thus, our result is a strong support of similarity between MERA and AdS/CFT correspondece.

The most recent development of the MERA is to construct the branching MERA network that reproduces anomalous entanglement-entropy scaling~\cite{Evenbly2}. The network looks like asymptotically the AdS space, but the IR geometry is separated into various types of multiple branches. In this case, the global structure of the MERA - tilde MERA combined network is topologically different from the present case. It will be an intersting work to examine how the genus affects the excitation modes on the conformal boundary.

\section{Discussion}

Based on the present results, we give some insights for black hole thermodynamics and for numerical renormalization group in condensed matter physics studied with use of TFD or dual CFT.

It has been pointed out that the Kruskal transformation in general relativity is similar to the Bogoliubov one in TFD~\cite{Israel,Maldacena2,Cantcheff,Czech}. Both of them use hyperbolic functions for the basis transformation. When we look at the Penrose diagram, the diagram separates the whole spacetime into four blocks. They are two subspaces, and black and white holes. In between two subspaces, the black hole appears. This seems to be quite similar to our combined network. When an observer is in one of two subspaces, it is considered that the observer can not contact with the other subspace. However, we think that the observer indirectly contacts with the other space through hidden quantum entanglement. This information corresponds to the black hole entropy.

The TFD has been also applied to the finite-temperature DMRG in quantum spin chains~\cite{Feiguin}. The natural ordering of sites for the best DMRG performance is to set up site $1$, ancilla $\tilde{1}$, site $2$, ancilla $\tilde{2}$, etc. The DMRG is a kind of diagonalization of the Hamiltonian with use of the reduced basis set. The basis set is determined by the eigenvectors with large eigenvalues of the reduced density matrix that is obtained by tracing over environmental degrees of freedom in the whole system. Then, we would like to reduce the quantum entanglement between near-neighbor sites as much as possible. Since the state $n$ and its ancilla $\tilde{n}$ are maximally entangled at high temperature, they should be paired in the simulation. This means that in the MPS optimized by the DMRG the ancilla blurs the pure-state properties on each site. On the other hand, in our combined MERA network, the bulk structure of the MERA network is modified. In that sense, the role of the tilde space on the coarse graining by finite temperature looks quite different in classical description (MERA) and direct quantum description (MPS). This may be a kind of duality between them.

\section{Summary}

Summarizing, we have shown that the dual tensor network formalism by the TFD is efficient for describing the finite-temperature properties of quantum systems. The TNS is a simple tensor product, and we do not assume any gravitational setup. However, the black hole thermodynamics  is automatically emerged from the formalism. In particular, the MERA network would be really representing the discrete AdS/CFT correspondence, and then it seems that Eq.~(\ref{cond2}) is a very strong constraint. In 2D maximally symmetric spaces, the negative curvature space only matches with the CFT in a group theoretical viewpoint, since the space is scale-invariant. Even if our MERA space is discretized, we think that the matching still remains. In the previous theories associated with MERA and AdS/CFT, the renormalization flow mechanism was mainly examined. On the other hand, the present study has approached the global spacetime structure of our extended MERA network. Thus, we hope that their interaction will leads to better understanding of MERA and AdS/CFT.

\begin{acknowledgements}
The work of M. I. was supported by World Premier International Research Center Initiative (WPI), MEXT, Japan.
\end{acknowledgements}

\end{document}